\documentclass[epj]{svjour}
\usepackage{epsfig}
\begin{document}
\title{ 
  On the  $\Delta$I=1/2 rule 
  in the
$\Lambda$N$\rightarrow$NN reaction
      }
\titlerunning{On the $\Delta$I=1/2 rule ...}
\authorrunning{Rudy, Z.}
\abstract{
  It is shown that the mass dependence of the $\Lambda$-lifetime in heavy
  hypernuclei is sensitive to the ratio of neutron-induced to
  proton-induced non-mesonic decay rates R$_n$/R$_p$.  A comparison of 
  the experimental mass dependence of the lifetimes with the calculated
  ones for different values of R$_n$/R$_p$ leads to the conclusion
  that this ratio is larger than 2 on the confidence level of 0.75.
  This suggests that the phenomenological  $\Delta$I=1/2 
  rule might be violated for the 
  nonmesonic decay of the $\Lambda$-hyperon.
         }
\author{
Z. Rudy$^{(a)}$, 
W. Cassing$^{(b)}$, 
L. Jarczyk$^{(a)}$, 
B. Kamys$^{(a)}$, 
P. Kulessa$^{(a,c)}$,
O. W. B. Schult$^{(c)}$,
A. Strza{\l}kowski$^{(a)}$
   }
\institute{
$^{(a)}$ M. Smoluchowski Institute of Physics, Jagellonian University, 
PL--30059 Cracow, Poland, \\ 
$^{(b)}$ Institut f\"ur Theoretische Physik, Universit\"at Giessen, 
D--35392 Giessen, Germany \\ 
$^{(c)}$ Institut f\"ur Kernphysik, Forschungszentrum J\"ulich, 
D--52425 J\"ulich, Germany\\ 
     }
\date{  }
\maketitle
%
%
%
\newcommand{\beq}{\begin{equation}}
\newcommand{\eeq}{\end{equation}}
\newcommand{\La}{$\Lambda$}
%
%
%
%
%
%
%
%

It is well known that weak, pionic decays of strange mesons and hyperons
strongly favour $\Delta$I=1/2 amplitudes over {\it a priori} comparable
$\Delta$I=3/2 amplitudes \cite{DAL63}.  This phenomenological rule may 
be interpreted on the level of a simple quark model as 
an indication that decays proceed via a transformation of the 
s-quark (with S=-1,I=0) into a u-quark (with S=0,I=1/2)
whereas all other quarks play the role of spectators.  The question arises 
whether the same, simple picture also holds in strangeness changing,
weak interaction between two baryons, such as a $\Lambda$-hyperon and a
nucleon. Since it is difficult to realize hyperon - nucleon scattering 
experimentally, the only practical way to study the weak YN interaction is
to examine the non-mesonic decays $\Lambda$p$\rightarrow$np and 
$\Lambda$n$\rightarrow$nn of hyperons bound in hypernuclei.
     The current theoretical models of the YN weak interaction, which rely
on the validity of the $\Delta$I=1/2 rule, have achieved 
a reasonable agreement
with data on the total non-mesonic decay rates but they are not able
to reproduce ratios of partial decay rates, i.e. neutron-induced to 
proton-induced decays \cite{COH90}.  Thus the applicability of the 
$\Delta$I=1/2 rule to the non-mesonic decay mode is an open question.
The analysis of existing data on non-mesonic decays of the lightest hypernuclei
\cite{SCH92} suggests that the $\Delta$I=1/2 rule might be violated in the weak 
hyperon-nucleon interaction.  It was also recently shown \cite{RAM97} that the 
theoretical analysis of the shape of proton spectra from the decay of 
$^{12}_{\Lambda}$C leads to the conclusion that the $\Delta$I=1/2 rule is not 
preserved in the non-mesonic decay.

The present letter examines another method of testing the validity of the 
$\Delta$I=1/2 rule in the non-mesonic decay.  It is based on the fact  that the 
mass dependence of the lifetime of hypernuclei is sensitive (for heavy 
hypernuclei) to the ratio R$_n$/R$_p$ of the neutron - induced (R$_n$) 
to proton - induced (R$_p$) decay rates of the $\Lambda$-hyperon.
Furthermore, it is known that the R$_n$/R$_p$ ratio should
be less or equal to 2 if the $\Delta$I=1/2 rule 
holds for the non-mesonic decay \cite{DAL63}. Thus an experimental evidence
for R$_n$/R$_p > 2$ obtained from the investigation of the 
mass dependence of the $\Lambda$- lifetime in heavy hypernuclei 
would imply a violation of the  $\Delta$I=1/2 rule.

We recall that the mass dependence of the $\Lambda$-lifetime in hypernuclei 
may be due to several effects:
\vspace{-0.4cm}
\begin{enumerate}
\item[i]  The Pauli blocking strongly affects the mesonic decay mode 
($\Lambda \rightarrow N \pi$) and depends on the mass of the hypernuclei,
but practically does not influence the nonmesonic decay process.  
This is caused by the different energy release in
mesonic ($\approx$ 40 MeV) and non-mesonic ($\approx$ 180 MeV) decays. 
Furthermore, this energy is almost equally shared among  both nucleons 
emerging from the non-mesonic decay, whereas the pion - due to momentum 
conservation - carries most of the energy released in the mesonic decay. 
Thus the nucleon in the final state from the mesonic decay is coming from far 
below the Fermi energy  whereas the final nucleons from the non-mesonic decay
are both highly above the Fermi energy.
\item[ii] Another source for the lifetime dependence on the mass of hypernuclei 
is the variation of the hyperon - nucleus potential with  
mass A of hypernuclei and a resulting variation of the  hyperon
wave function. This dependence in turn will affect both, the mesonic and  
non-mesonic decay rates.  
\item[iii] The last effect 
is the variation of the ratio N/Z with the mass of hypernuclei.
This is the most important
effect for our purposes because such a variation can lead to
a mass dependence of the $\Lambda$-lifetime in hypernuclei  only
if the ratio R$_n$/R$_p$ deviates from unity. Thus the mass dependence
of the lifetime of heavy hypernuclei appears to be sensitive to
the R$_n$/R$_p$ ratio.
\end{enumerate}

In order to estimate the decay width of the mesonic decay a phase-space model 
has been  used in line with the Coupled - Channel - Boltzmann - Uhling - 
Uhlenbeck approach \cite{RUD95} .  The decay width in units of the free 
$\Lambda$-hyperon decay width including Pauli blocking is given by:
  $$
  \Gamma_{mesonic}/\Gamma_{free} =
  \frac{1}{(2 \pi)^3} \int d^3 r \int d^3 p
  \int d^3 p' 
  $$
\begin{equation}
  \label{formula4}
f_{\Lambda}({\bf r, p})
(1 - f_N({\bf r, p'}))
\delta({\bf p-p_\pi-p'}) 
\end{equation}
with $p_\pi = 102 MeV/c$,
where $f_{\Lambda}({\bf r, p})$ is approximated by a Gaussian s-state
phase-space density from the harmonic oscillator model
\begin{eqnarray}
f_{\Lambda}({\bf r, p})=\frac{1}{\pi^3}
exp(-\frac{\bf r^2}{\gamma^2}) \times
exp(-\gamma^2{\bf p^2})   \hspace*{1.0cm} ;  \nonumber \\
  \gamma=\sqrt{\frac{\hbar}{\mu \omega}} \hspace{0.6cm} ; \hspace{0.6cm}
\hbar\omega = 41 A^{-1/3} [MeV]
\end{eqnarray}
normalized to unity, whereas the nucleon
phase-space distribution is taken in the semiclassical limit, i.e.
\newline 
\begin{equation}
 f_N({\bf r, p'}) = \Theta(p_F({\bf r})-p')  
\end{equation}

\noindent
with $p_F({\bf r})$ denoting the local
Fermi momentum defined by the experimental density for the nucleus
of interest, $ p_F({\bf r}) = \left(\frac{3}{2}\pi^2 \rho({\bf r})
 \right)^{1/3}$ and $\Theta(x)$ is the step - function, i.e.
 $\Theta(x)$=1 for $x \ge 0$ and $\Theta(x)$=0 for $x<0$.

The same conceptional approach is applied to the non-mesonic decay and  
leads to the following width that also includes the blocking
of the final nucleon states \cite{RUD95}:
$$
\Gamma_{non-mesonic} \approx R_{N}  \frac{4}{(2\pi)^6}
\int d^3 r \int d^3 p \int d^3 p_N \int d^3 p_1'
$$
  $$
  \int \frac{d \Omega}{4 \pi}
  \ | v_{\Lambda N} |  
  \frac{d \sigma_{\Lambda N \rightarrow \Lambda N}}
  {d \Omega} (\sqrt{s}) 
  f_{\Lambda}({\bf r, p}) f_N({\bf r, p_N}) \times
  $$
\begin{equation}  \label{alphapar}
(1- f_N({\bf r, p_2'})
(1 - f_N({\bf r, p_1'}) \delta({\bf p+p_N-p_1'-p_2'}), \hspace{0.5cm}
\end{equation}
\noindent
where the phase-space distributions are the same as in eq. 
(\ref{formula4}) and the
differential cross section $d \sigma_{\Lambda N \rightarrow N N}/d \Omega$
has been approximated by $R_{N}  \ d\sigma_{\Lambda N \rightarrow \Lambda N}
/d \Omega$ \cite{RUD95} 
with a weak decay constant $R_{N}$. The latter is the squared ratio of the
weak transition matrix element to that for the strong interaction 
 $\Lambda$N$\rightarrow \Lambda$N.
The quantity $v_{\Lambda N}$, furthermore, is the relative velocity of  
the $\Lambda$-hyperon and the nucleon in their collision.
The constant $R_{n}$ for the process 
$n + \Lambda \rightarrow n + n$ might have a different value  
from the constant  $R_{p}$
for the $p + \Lambda \rightarrow p + n$ channel.
This is due to the fact that the final NN system can have
both isospin $I_{NN}=0$  or $I_{NN}=1$ in case of neutron--proton
final states and  only  $I_{NN}=1$  in case of two final neutrons.
Anyway, one can always write 
$\Gamma_{non-mesonic}$ = $\Gamma_n$ + $\Gamma_p$, where
$\Gamma_n$ and $\Gamma_p$ can be separately calculated via formula 
(\ref{alphapar}) using instead of 
R$_N$ the effective strength equal to N$\cdot$R$_n$/A or Z$\cdot$R$_p$/A for 
neutrons and protons, respectively.

   \renewcommand{\topfraction}{0.90}
\begin{figure}[h]
\begin{minipage}{15cm}
\epsfig{file=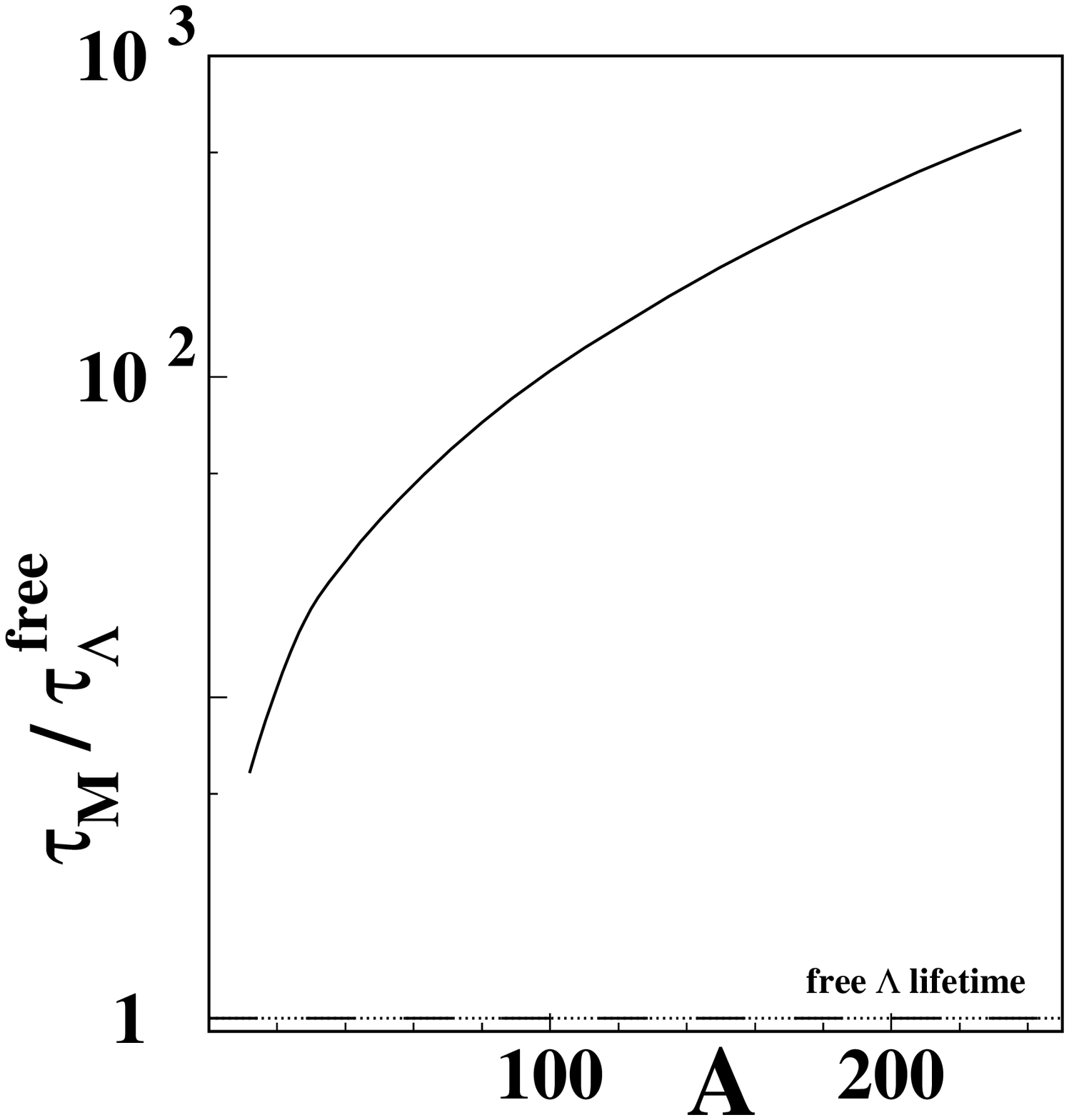,height=60mm,
   ,angle=0,bbllx=20,bblly=140,bburx=580,bbury=640}
\end{minipage}
\begin{minipage}{15cm}
\epsfig{file=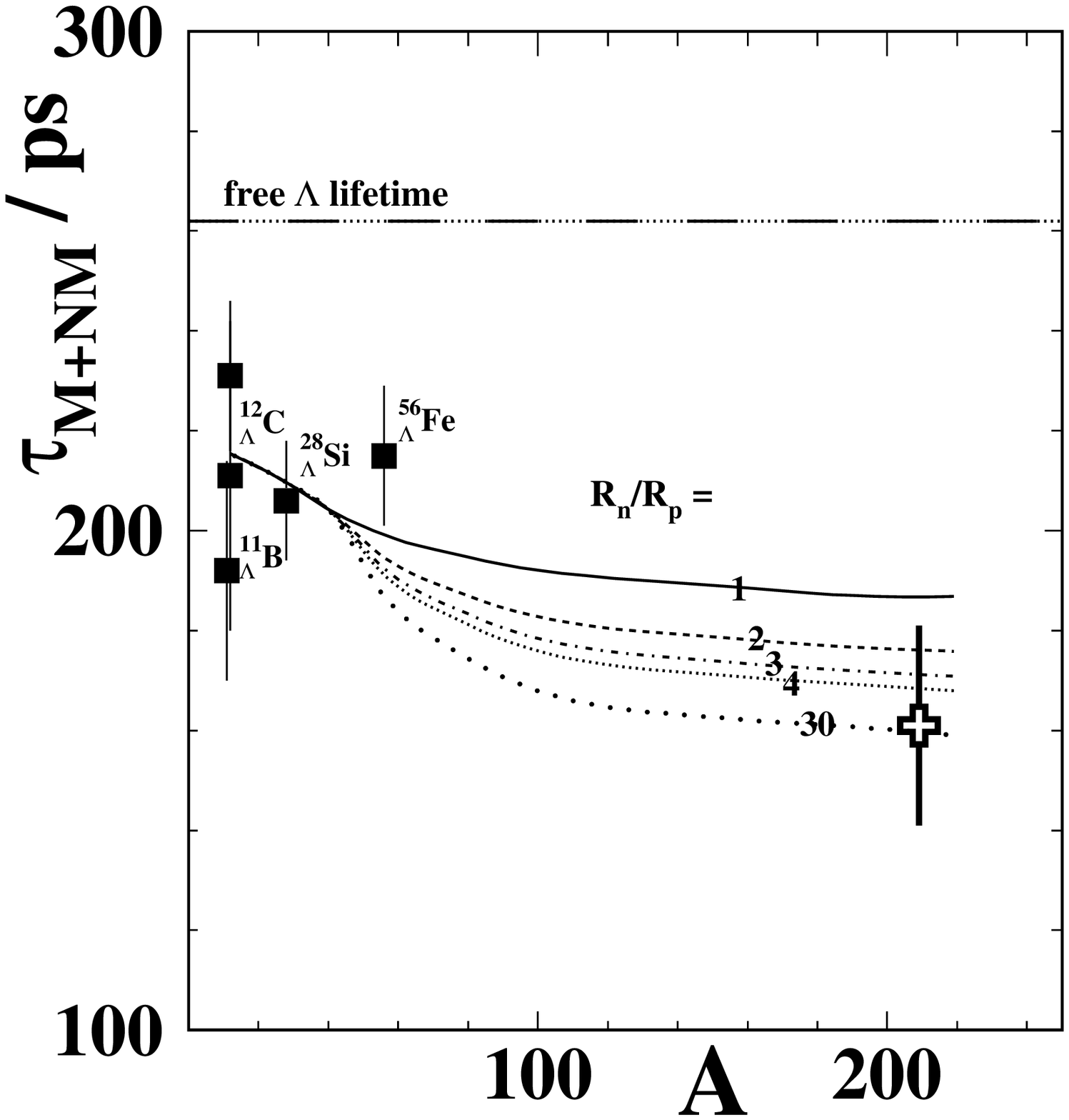,height=60mm,
        ,angle=0,bbllx=20,bblly=140,bburx=580,bbury=640}
\end{minipage}
{\caption{ \footnotesize \label{pauliblo}
 Upper part: phase--space model calculation of the lifetime  $\tau_M$ 
of the $\Lambda$--hyperon
due to the mesonic decay  inside the hypernucleus of mass A  
according to eq.  (\ref{formula4}) presented in units of the lifetime of the 
free $\Lambda$--hyperon $\tau^{free}_{\Lambda}$ = 263 ps; 
 Lower part:
 calculations of the $\Lambda$-lifetime $\tau_{M+NM}$ due to
the mesonic and nonmesonic decay  as a function 
 of hypernucleus mass A in comparison to the data of 
Ref. \cite{BHA98,SZY91,KUL98}. 
 The COSY--13 collaboration result for a  Bi target \cite{KUL98}
 is marked by the cross.
 In the theoretical calculations both mesonic and  non-mesonic decay modes  
 are taken into account whereas 
 the  unknown ratio of the weak decay rates
   $R_{n}$/$R_{p}$ in equation (\ref{alphapar})
 is varied assuming the values:   $R_{n}$/$R_{p}$ =  1,2,3,4 and 30.
}        }
\end{figure}

The  lifetime $\tau_{M} = \hbar/\Gamma_{mesonic}$ of the 
$\Lambda$--hyperon due to the pure $\pi N$ decay channel
evaluated according to formula (\ref{formula4})
is displayed in the upper part of Fig. \ref{pauliblo}
as a function of the mass A of the hypernucleus.
It shows a dramatic
increase of $\tau_{M}$ due to Pauli-blocking
what contradicts the tendency of the experimental
mass dependence of the lifetime
depicted in the lower part of Fig. \ref{pauliblo}. 
This indicates clearly that the non--mesonic decay
dominates for medium--mass and heavy hypernuclei.

The results for $\tau_{\Lambda} = \hbar/(\Gamma_{mes} + \Gamma_{non-mesonic})$,
where $\Gamma_{non-mesonic}$ = $\Gamma_{n}$ + $\Gamma_{p}$
are shown in the lower part of Fig. \ref{pauliblo} as a function 
of the mass A.
The calculations were performed for several ratios
$R_n$/$R_p$ while keeping 
$R_{av}$ = $(R_{n}$ + $R_{p})$/2
constant. $R_{av}$ was 
fixed by the requirement that the constructed model
has to describe the lifetime data for 
light hypernuclei, i.e. $^{11}_\Lambda$B and  
$^{12}_\Lambda$C \cite{BHA98,SZY91}.

The lifetimes calculated 
for hypernuclei in the neighbourhood of mass number A=200 are 
equal to $\approx$175 ps for the ratio $R_{n}$/$R_{p}$=2, i.e. the 
limiting value which is still compatible with the $\Delta$I=1/2 rule.  
Smaller values of the lifetime for these heavy hypernuclei 
correspond to ratios $R_{n}$/$R_{p} > 2 $ and therefore
they imply a violation of the $\Delta$I=1/2 rule. 
It should be noted, that in order to obtain a convincing conclusion 
about the violation of the $\Delta$I=1/2 rule the errors of 
the lifetime measurements should be very small.

Recently, the COSY--13 collaboration \cite{KUL98} has 
measured the lifetime for heavy hypernuclei
produced in the p+Bi interaction with experimental
errors distinctly smaller than other  published results \cite{ARM93,OHM97}.
This value for the lifetime $\tau_{\Lambda}$ = (161$\pm$21) ps 
leads to the conclusion that the 
R$_n$/R$_p$ ratio is larger than 2 on the confidence level
of 0.75. This suggests that the phenomenological $\Delta$I=1/2 
rule may be violated for the non--mesonic decay  of the $\Lambda$--hyperon.
It should be noted that the lines
displayed in Fig. 1 have been obtained for the average N/Z ratio
as expected \cite{RUD96} for the cold hypernuclei produced
in the p+Bi experiment \cite{KUL98}.

The estimation of the confidence 
level was performed taking into account both, the inaccuracy of 
the R$_{av}$
due to the experimental errors for the lifetime of light hypernuclei and 
the inaccuracy of the experimental determination of the lifetimes for heavy
hypernuclei.
It was assumed that the distribution of R$_{av}$ as well as
the distribution of the experimental lifetime of heavy hypernuclei 
are described
by normal distributions, i.e.  N(${<}R_{av}{>},\sigma(R_{av}$)) and 
N(${<}\tau_{\Lambda}{>},\sigma(\tau_{\Lambda})$), respectively. 
The parameters
${<}R_{av}{>}$, $\sigma(R_{av}$) were found from a fit of the theoretical
curve to the lifetimes of the light hypernuclei, i.e. $^{11}_{\Lambda}$B
and  
$^{12}_{\Lambda}$C \cite{BHA98,SZY91}
%
%
whereas ${<}\tau_{\Lambda}{>}$ and $\sigma(\tau_{\Lambda})$ are
known from the p+Bi experiment \cite{KUL98}.
%
Then the confidence level has been evaluated by convolution of these
two distributions according to:
\begin{eqnarray}
P & = &{\int}_{-\infty}^{+\infty} dR_{av} \  N({<}R_{av}{>},\sigma(R_{av})) \cdot \\
\nonumber
  &   & \cdot \int_{-\infty}^{\tau_{max}(R_{av},\frac{R_{n}}{R_{p}}=2)}
d\tau_{\Lambda}  \  N({<}\tau_{\Lambda}{>},\sigma(\tau_{\Lambda})) , 
\end{eqnarray}
where $\tau_{max}(R_{av},\frac{R_{n}}{R_{p}}=2)$ is the theoretical
lifetime evaluated for heavy hypernuclei with the proper value of R$_{av}$ and
limiting value of the R$_n$/R$_p$ ratio.

\newpage
In order to put more stringent limits
to the $R_{n}$/$R_{p}$ ratio and thus
to test further the validity of the $\Delta$I=1/2 rule
on a higher confidence level
it is necessary to obtain new precise data
on the lifetime
of medium--mass and heavy hypernuclei.
%
%
Such experiments
are planned both by the COSY--13 collaboration
at Forschungszentrum J\"ulich \cite{BOR98}
for heavy hypernuclei and by the E369 collaboration
at KEK for $_{\Lambda}^{89}$Y \cite{BHA98}.

  It should be emphasized, that the origin 
of the  $\Delta$I=1/2 rule is not fully understood 
at present \cite{PAR98}, though theoretical calculations in most cases
rely on the validity of this rule \cite{COH90}.
Therefore, experimental results which indicate a violation of this rule will  
have relevant implications for all theoretical
models of the weak interaction 
of baryons.  \vspace*{0.3cm} \\

{\footnotesize This work was partly supported by the International Bureau of the BMBF,
Bonn, DLR, and by the Polish Committee for Scientific Research.}
\end{document}